\documentclass[conference,10pt]{IEEEtran}

\usepackage[final]{graphicx}
\usepackage{graphics}
\usepackage{amsmath,amsthm,amssymb,algorithmic}
\usepackage[margin=1in]{geometry}
\usepackage{subfig}
\usepackage{color}
\usepackage{setspace}
\usepackage{hyperref}
\usepackage{epstopdf}
\usepackage{soul}
\hypersetup{colorlinks=true}

 \hypersetup{colorlinks=true}
\newcommand{\ben}{\begin{eqnarray}}

\newcommand{\een}{\end{eqnarray}}

\newcommand{\sign}{\mathop{\bf sgn}}
\newcommand{\transpose}{^{\top}}
\newtheorem{thm}{Theorem}
\newtheorem{corollary}{Corollary}

\newcommand{\sgn}{\textbf{sgn}}

\newcommand{\y}{\textbf{y}}

\newcommand{\ab}{\textbf{a}}
\newcommand{\bb}{\textbf{b}}

\newcommand{\xb}{\textbf{x}}

\newcommand{\A}{\textbf{A}}

\newcommand{\G}{\textbf{G}}
\newcommand{\R}{\textbf{R}}

\newcommand{\X}{\textbf{X}}
\newcommand{\I}{\textbf{I}}

\newcommand{\yb}{\textbf{y}}

\newcommand{\vb}{\textbf{v}}
\newcommand{\wb}{\textbf{w}}
\newcommand{\la}{\langle}
\newcommand{\ra}{\rangle}

\begin{document}
%
\title{Reduced-Dimension Multiuser Detection}

\author{\IEEEauthorblockN{Yao Xie}
\IEEEauthorblockA{Dept. of Electrical Engineering\\
Stanford University\\
Email: yaoxie@stanford.edu}
\and
\IEEEauthorblockN{Yonina C. Eldar}
\IEEEauthorblockA{Dept. of Electrical Engineering\\
Technion, Isreal Institute of Technology\\
Email: yonina@ee.technion.ac.il}
\and
\IEEEauthorblockN{Andrea J. Goldsmith}
\IEEEauthorblockA{Dept. of Electrical Engineering\\
Stanford University\\
Email: andrea@wsl.stanford.edu}}

\maketitle

\begin{abstract}

   We explore several reduced-dimension multiuser detection (RD-MUD) structures that significantly decrease the number of required correlation branches at the receiver front-end, while still achieving performance similar to that of the conventional matched-filter (MF) bank. RD-MUD exploits the fact that the number of active users is typically small relative to the total number of users in the system and relies on ideas of analog compressed sensing to reduce the number of correlators. We first develop a general framework for both linear and nonlinear RD-MUD detectors. We then present theoretical performance analysis for two specific detectors: the linear reduced-dimension decorrelating (RDD) detector, which combines subspace projection and thresholding to determine active users and sign detection for data recovery, and the nonlinear reduced-dimension decision-feedback (RDDF) detector, which combines decision-feedback orthogonal matching pursuit for active user detection and  sign detection for data recovery. The theoretical performance results for both detectors are validated via numerical simulations. \let\thefootnote\relax\footnotetext{This work is partially supported by the Interconnect Focus Center of the Semiconductor Research Corporation, BSF Transformative Science Grant 2010505, and a Stanford General Yao-Wu Wang Graduate Fellowship. 
Submitted to IEEE International Conference on Communications (ICC) 2012. Copyright IEEE. 
}
    
\end{abstract}


%
\IEEEpeerreviewmaketitle

\section{Introduction}\label{sec:intro}

Multiuser detection (MUD) \cite{verduMUD1998} is a classical problem in multiuser communications, where a number of users communicate simultaneously with a given receiver by modulating information symbols onto their unique signature waveforms. The received signal consists of a noisy version of the superposition of the transmitted waveforms, and the receiver has to detect the symbols of all users simultaneously.
While there has been a large body of work developed for the multiuser detection problem, it is not yet widely implemented in practice, largely due to its complexity and high-precision A/D requirement. The complexity of MUD arises both in the analog circuitry for decorrelation as well the digital signal processing for data detection of each user. We characterize the decorrelation complexity by the number of correlators used and the data detection complexity by the complexity-per-bit \cite{verduMUD1998}.

The conventional MUD detection structure consists of a matched-filter (MF) bank front-end followed by a linear or nonlinear digital detector. The MF-bank front-end is a set of correlators, each correlating the received signal with the signature waveform of a different user. Hence the conventional MUD requires the number of correlators to be equal to the number of users. To recover user data from the MF-bank output, various digital detectors have been developed. The optimal MUD is the maximum likelihood sequence estimator (MLSE) \cite{verduMUD1998}, which minimizes the probability of error for symbol detection, but its complexity-per-bit is exponential in the number of users when the signature waveforms are nonorthogonal. The nonlinear decision feedback (DF) detector \cite{verduMUD1998} is a good compromise between complexity and performance among all nonlinear and linear MUD techniques \cite{verduMUD1998}. This technique detects users iteratively and subtracts the strongest user in each iteration. Both the MLSE and the DF detectors are nonlinear methods. 
Linear detection requires lower complexity but with a commensurate reduction in performance. This technique applies a linear transform to the receiver front-end output and then detects each symbol separately. Linear MUD techniques include the single-user detector, the decorrelating detector and the minimum mean-square-error (MMSE) detector \cite{verduMUD1998}. When the user signature waveforms are correlated, the performance of the single-user detector degrades, while the decorrelating detector \cite{verduMUD1998} eliminates the user interference by projecting the received signal onto the subspace of the signature waveform of each user. The decorrelating detector optimizes the near-far resistance among linear detectors \cite{verduMUD1998}, although it also amplifies noise.  
%
Both linear and nonlinear MUDs have sufficiently high complexity to preclude their wide adoption in deployed systems. One reason is that they both require the number of correlators at the receiver front-end to be equal to the number of users in the system. 

In an earlier work \cite{XieEldarGoldsmith2010RDMUD}, we introduced the structure of a low complexity reduced-dimension multiuser detection (RD-MUD). The RD-MUD exploits the fact that the number of active users $K$ is typically much smaller than the total number of users $N$ at any given time. 
Our RD-MUD has a front-end that correlates the received signal with $M$ correlating signals, with $M$ much smaller than $N$. The correlating signals are formed as linear combinations of the signature waveforms via a (possibly complex) coefficient matrix $\A$. Our choice of $\A$ will be shown to be crucial for performance. The output of the RD-MUD front-end can thus be viewed as a projection of the MF-bank output onto a lower dimensional \textit{detection subspace}. 

After first developing structures for general linear and nonlinear RD-MUDs, 
we will develop performance analysis bounds for two of these structures: the reduced-dimension decorrelating (RDD) detector, a linear detector that combines subspace projection and thresholding to determine active users with a sign detector for data recovery \cite{BlumensathDavies2009}, and the reduced-dimension decision-feedback (RDDF), a nonlinear detector that combines decision-feedback orthogonal matching pursuit (DF-OMP) \cite{Tropp2004} for active user detection with the sign detector for data recovery in an iterative manner. We present theoretical probability-of-error performance guarantees for these two detectors in terms of the coherence of the matrix $\A$, in a non-asymptotic regime with a fixed number of users and active users.  Our RD-MUD detectors consists of two stages: active user detection and data detection of active users. The first stage is closely related to \cite{Ben-HaimEldarElad2010}. However, our problem differs in that the probability-of-error must consider errors in both stages. We derive conditions under which the probability-of-error is dominated by errors in the first stage. 
We do not consider optimizing signature waveforms and hence our results will be parameterized by the crosscorrelation properties of the given signature waveforms. 

The rest of the paper is organized as follows. Section \ref{sec:model} and Section \ref{sec:RD-FE} present the model and the RD-MUD front-end, respectively. Section \ref{sec:detectors} introduces the digital detectors we propose for RD-MUD. Section \ref{sec:RD-MUD-performance} contains the theoretical performance guarantees. Section \ref{sec:numerical_eg} contains numerical examples, and finally Section \ref{sec:conclusion} concludes the paper.

\section{System Model}\label{sec:model}

Consider a multiuser system with $N$ users. Each user is assigned a unique signature waveform from a set $\mathcal{S} = \{s_n(\cdot): [0, T]\rightarrow \mathbb{R}, 1 \leq n \leq N\}$, which are assumed given and known, and posess certain properties discussed in more detail below. 
Each user modulates its signature waveform to transmit its symbols. The symbols carry information. The duration of the signature waveforms $T$ is referred to as the symbol time. 
Define the \textit{inner product} (or \textit{crosscorrelation}) between two real analog signals $x(t)$ and $y(t)$ as $\la x(t), y(t)\ra \triangleq T^{-1}\int_{0}^T x(t) y(t) dt$. 
%
The crosscorrelations of the signature waveforms are characterized by the Gram matrix $\G$, defined as
\begin{equation}
[\G]_{nl}\triangleq \la s_n(t), s_l(t) \ra, \quad 1\leq n \leq N, \quad 1\leq l\leq N.
\end{equation}
For convenience, we assume that $s_n(t)$ has unit energy: $\|s_n(t)\|^2 \triangleq \la s_n(t), s_n(t)\ra = 1$ for all $n$ so that $[\G]_{nn} = 1$. We also assume that the signature waveforms are linearly independent. Hence $\G$ is invertible. We consider the synchronous MUD model that uses Binary Phase Shift Keying (BPSK) modulation \cite{verduMUD1998}. There are $K$ active users with index set $n \in \mathcal{I}$. The complement set $\mathcal{I}^c$ contains indices of all non-active users.
The symbol of the user $n$ is $b_n \in \{1, -1\}$, for $n\in\mathcal{I}$. Define a gain factor $r_n$ for each user which captures the transmitting power and channel gain. We assume $r_n$ is real and known to the receiver. The nonactive users can be viewed as transmitting with zero power, or equivalently transmitting zeros: $b_n = 0$, for $n\in\mathcal{I}^c$.
%
%
The received signal $y(t)$ is a superposition of the transmitted signals from the active users, plus white Gaussian noise $w(t)$ with zero-mean and variance $\sigma^2$:
\begin{equation}
    y(t) =  \sum_{n=1}^N r_n b_n s_n(t) + w(t), \qquad t \in [0, T], \label{sig_model}
\end{equation}
with $b_n \in \{1, -1\}$, $n\in\mathcal{I}$, and $b_n = 0$, $n \in \mathcal{I}^c$.
%
The goal of multiuser detection (MUD) is to detect the set of active users $\mathcal{I}$ and their transmitted symbols $\{b_n: n\in\mathcal{I}\}$. In practice the number of active users $K$ is typically much smaller than the total number of users $N$, which is a form of \textit{user sparsity}. Therefore, the received signal $y(t)$ consists of only a few transmissions from active users. As we show, this user sparsity enables us to reduce the number of correlators at the front-end and still be able to achieve performance similar to that of a conventional MUD using a bank of MFs. To simplify the detection algorithm, we assume that $K$ is known. 
The problem of estimating $K$ can be treated separately \cite{BiglieriLops2007}.

\vspace{0.15in}
\begin{figure}[h]
    \centering
          \includegraphics[width=2in]{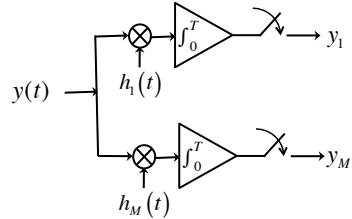}
        \caption{Front-end of RD-MUD.}
        \label{Fig:front_end}
    \end{figure}
    \vspace{-0.1in}

\section{RD-MUD Front-End}\label{sec:RD-FE}

\begin{figure}[h]
    \centering
        \includegraphics[width=3in]{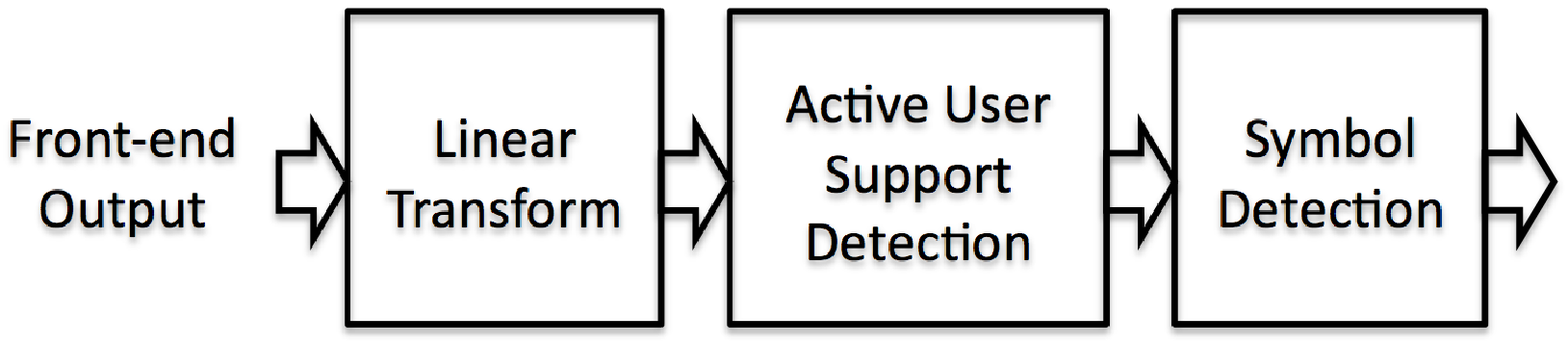}
          \includegraphics[width=2in]{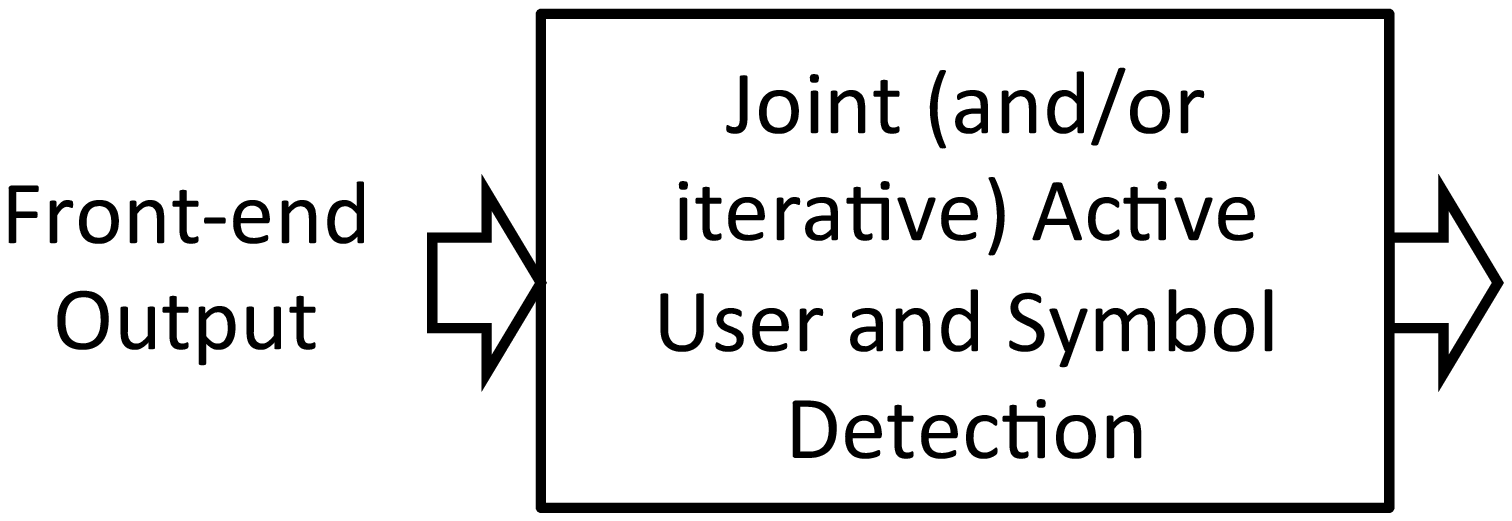}
        \caption{The diagram of (upper) linear detector, and (lower) nonlinear detector.}
        \label{Fig:detector}
    \end{figure}

The RD-MUD front-end, illustrated in Fig. \ref{Fig:front_end}, correlates the received signal $y(t)$ with a set of correlating signals $h_m(t)$, $m = 1, \cdots M$, where $M$ is typically much smaller than $N$. This is in contrast to the conventional matched filter (MF) bank, which correlates the received signal with the full set of $N$ signature waveforms \cite{verduMUD1998}. 
The front-end output is processed by either a linear or nonlinear detector to detect active users and their symbols, as shown in Fig. \ref{Fig:detector} for both linear and nonlinear detectors. 
The design of the correlating signals $h_m(t)$ is the key for RD-MUD to reduce the number of correlators. To construct these signals, we rely on biorthogonal waveforms \cite{Ben-HaimEldarElad2010}. 
The biorthogonal signals with respect to $\{s_n(t)\}$ are defined as a linear combination of all signature waveforms using a weighting coefficient matrix $\G^{-1}$: $
\hat{s}_n(t) = \sum_{l=1}^N  [\G^{-1}]_{nl} s_l(t)$,  $1\leq n \leq N$, where $[\X]_{nm}$ denotes the element of a matrix $\X$ at its $n$th row and the $m$th column. Also denote by $[\xb]_n$ the $n$th entry of a vector $\xb$. The biorthogonal signals have the property that $\la s_n(t), \hat{s}_m(t) \ra = \delta_{n, m}$, for all $n$, $m$. The delta function $\delta_{n, m}$ is equal to one when $n = m$, and is equal to zero otherwise. The correlating signals $h_m(t)$ are linear combinations of the biorthogonal waveforms with  (possibly complex) weighting coefficients $a_{mn}$ that we choose:
\begin{equation}
h_m(t) = \sum_{n = 1}^{N} a_{mn} \hat{s}_n(t), \qquad 1\leq m\leq M. \label{h_def}
\end{equation}
Define a coefficient matrix $\A \in \mathbb{R}^{M\times N}$ with $[\A]_{mn} \triangleq a_{mn}$ and denote the $n$th column of $\A$ as $\ab_n \triangleq [a_{1n}, \cdots, a_{Mn}]\transpose$, $n = 1, \cdots, N$. The notation $\X\transpose$ denotes the transpose of a vector or matrix. We normalize the columns of $\A$ so that $\|\ab_n\|^2 \triangleq \sum_{m=1}^M a_{nm}^* a_{nm}= 1$, where $x^*$ is the conjugate of a scalar $x$.
The design of the correlating signals is equivalent to the design of the coefficient matrix $\A$ for a given $\{s_n(t)\}$. We will use \textit{coherence} as a measure of the quality of $\A$, which is defined as \cite{Ben-HaimEldarElad2010}:
\begin{equation}
\mu\triangleq \max_{n\neq l}\left|\ab_n^H \ab_l\right|.\label{def_coherence}
\end{equation}
As we will show later, it is desirable that the columns of $\A$ have small correlation such that $\mu$ is small. The output of the $m$th correlator is given by $y_{m}  = \la h_m(t), y(t) \ra$.  Denoting $\yb = [y_{1}, \cdots, y_{M}]\transpose$, we can derive the output of the RD-MUD front-end as (detailed derivations can be found in \cite{XieEldarGoldsmith2010RDMUD}):
\begin{equation}
    \yb = \A \R\bb + {\wb},\label{RD_MUD_model}
\end{equation}
where $\wb$ is a Gaussian random vector with zero mean and covariance $\sigma^2 \A\G^{-1}\A^H$, $\textbf{R}$ is a diagonal matrix with $r_{nn}$ on the diagonal, and $\bb\triangleq  [b_1, \cdots, b_N]\transpose$. The notation $\X^H$ denotes the conjugate transpose of a matrix $\X$. The vector $\yb$ can be viewed as a linear projection of the MF-bank front-end output onto a lower dimensional subspace which we call the \textit{detection subspace}. Since there are at most $K$ active users, $\bb$ has at most $K$ non-zero entries. The idea of RD-MUD is that when the original signal vector $\bb$ is sparse, with proper choice of the matrix $\A$, the detection performance for $\bb$ based on $\yb$ of (\ref{RD_MUD_model}) in the detection subspace can be similar to the performance based on the output of the MF-bank front-end.
    \vspace{0.15in}
    
\section{RD-MUD Detectors}\label{sec:detectors}

We now discuss how to recover $\bb$ from the RD-MUD front-end output $\yb$ of (\ref{RD_MUD_model}) using digital detectors. The model for the output (\ref{RD_MUD_model}) of the RD-MUD front-end has a similar form to the observation model in the compressed sensing literature \cite{FletcherRanganGoyal2009}\cite{Ben-HaimEldarElad2010}, except that the noise in the RD-MUD front-end output is colored due to match filtering at the front-end. Hence, to recover $\bb$, we can adopt the ideas developed in the context of compressed sensing, and combine them with techniques of MF-bank detection. 

The linear detector for RD-MUD first recovers active users $\hat{\mathcal{I}}$ using support recovery techniques from compressed sensing (e.g., \cite{FletcherRanganGoyal2009}).  Given an index set $\mathcal{I}$, $\X_{\mathcal{I}}$ denotes the submatrix formed by the columns of a matrix $\X$ indexed by $\mathcal{I}$, and  $\xb_{\mathcal{I}}$ denotes the subvector formed by the entries indexed by $\mathcal{I}$.
Based on the recovered index set of active users $\hat{\mathcal{I}}$, we can write the RD-MUD front-end output model (\ref{RD_MUD_model}) as
\begin{equation}
\yb = \A_{\hat{\mathcal{I}}}\R_{\hat{\mathcal{I}}}\bb_{\hat{\mathcal{I}}} + \wb.\label{restrict}
\end{equation}
Once the active users are detected, their symbols $\bb_{\hat{\mathcal{I}}}$ can be detected from (\ref{restrict}). This is done by applying a linear transform to the front-end output and detecting symbols separately. The nonlinear detector for RD-MUD detects  active users and their symbols jointly (and/or iteratively).  

We will focus on recovery based on two algorithms: (1) the reduced-dimension decorrelating (RDD) detector, a linear detector that uses subspace projection along with thresholding \cite{BlumensathDavies2009} to determine active users and sign detection for data recovery; (2) the reduced-dimension decision feedback (RDDF) detector, a nonlinear detector that combines decision-feedback orthogonal matching pursuit (DF-OMP) for active user detection and sign detection for data recovery. DF-OMP differs from the conventional OMP \cite{Tropp2004} in that in each iteration, the binary-valued detected symbols, rather than the real-valued estimates, are subtracted from the received signal to form the residual used by the next iteration. The residual consists of the remaining undetected active users. By subtracting interference from the strongest active user we make it easier to detect the remaining active users.

\subsection{Reduced-dimension decorrelating (RDD) detector}\label{sec:RDD}

The RDD detector works as follows. As per (\ref{RD_MUD_model}), the front-end of the RD-MUD projects the received signal $y(t)$ onto the detection subspace as a vector $\yb$. By considering the RD-MUD front-end output when the input signal is $s_n(t)$, we can show that the column $\ab_n$ of $\A$ corresponds to the $n$th signature waveform vector in the detection subspace. Considering the detection method of the conventional MUD, a natural strategy for RD-MUD is to match the received signal vector $\yb$ and the $n$th signature waveform vector in the detection subspace by computing their inner product, which is given by $\ab_n^H \yb$, $n = 1, \cdots, N$. To detect active users, we can rank the magnitudes of these inner products and detect the index of the $K$ largest as active users:
\begin{equation}
\begin{split}
&\hat{\mathcal{I}} = \{n: \quad \mbox{if $|\Re[\ab_n^H \yb]|$} \\
&\mbox{is among the $K$ largest of $|\Re[\ab_n^H\yb]|$, $n = 1, \cdots, N$} \},
\end{split}
\label{support}
\end{equation}
where $\Re[x]$ denotes the real part of a number $x$.
To detect their symbols, we use sign detection:
\begin{equation}
    \hat{b}_n = \left\{
    \begin{array}{cc}
     \sgn\left(r_n\Re[\ab_n^H \yb] \right), & n\in\hat{\mathcal{I}};\\
     0, &n\notin\hat{\mathcal{I}}.
     \end{array}\right.
    \label{RD-MUD-sign}
\end{equation}
where $\sgn(x)$ denotes the sign of a number $x$.
In detecting active users (\ref{support}) and their symbols (\ref{RD-MUD-sign}), we take real parts of the inner products because the imaginary part of $\ab_n^H \yb$ contains only noise and interference.
The complexity-per-bit for data detection of the RDD detector is proportional to $M$. Since $M\leq N$ in RD-MUD, the complexity for data detection of the RDD detector is on the same order as that of the conventional linear MUD detector. But the RDD detector requires much lower decorrelation complexity than the conventional linear detector. 

\subsection{Reduced-dimension decision feedback (RDDF) detector}\label{sec:algorithm_II}

The RDDF detector detects active users and symbols iteratively. It starts with an empty set as the initial estimate for the set of active user $\hat{\mathcal{I}}^0$, zeros as the estimated symbol vector $\bb^{(0)} = \textbf{0}$, and the front-end output as the residual vector $\vb^{(0)}=\y$. Subsequently, in each iteration $k = 1, \cdots, K$, the algorithm selects the column $\ab_n$ that is most highly correlated with the residual $\vb^{(k-1)}$ as the detected active user in the $k$th iteration, with the active user index: 
\begin{equation}
n_k=\arg\max_n \left|\Re[\ab_n^H \vb^{(k-1)}]\right|.
\end{equation} 
This index is then added to the active user set $\hat{\mathcal{I}}^{(k)} = \hat{\mathcal{I}}^{(k-1)}\cup\{n_k\}$.
The symbol for user $n_k$ is detected with other detected symbols staying the same:
\begin{equation}
{b}_{n}^{(k)} = \left\{
\begin{array}{cc} \sign(\Re[r_{n_k} \ab_{n_k}^H \vb^{(k-1)}]), & n = n_k;\\
{b}_n^{(k-1)}, & n \neq n_k.
\end{array}\right.
\end{equation}
Then the residual vector is updated through
\begin{equation}
\vb^{(k)} = \yb - \A\R\bb^{(k)}. 
\end{equation}
The iteration repeats $K$ times (we show in \cite{XieEldarGoldsmithJournal2011} that with high probability DF-OMP never detects the same active user twice), and finally the active user set is given by $\hat{\mathcal{I}}=\hat{\mathcal{I}}^{(K)}$ with the symbol vector $\hat{b}_n = b^{(K)}_n$, $n = 1, \cdots, N$. The complexity-per-bit of the RDDF detector is proportional to $MK$. Since $M\leq N$, this implies that the complexity for data detection of the RDDF detector is on the same order as that of the conventional DF detector. But the RDDF detector requires much lower decorrelation complexity than the conventional DF detector. 

\subsection{Reduced-Dimension MMSE (RD-MMSE) Detector}\label{sec:RD_MUD_decorrelator}

Similar to the MMSE detector of the conventional MUD, a linear detector based on the MMSE criterion can be derived for the reduced-dimension model (\ref{restrict}) (see \cite{XieEldarGoldsmithJournal2011} for derivations). The RD-MMSE detector detects the set of active users $\hat{\mathcal{I}}$ first by a support recovery method and then detects symbols as:
\begin{equation}
\hspace{-0.2in}\hat{b}_n =
\left\{\begin{array}{cc}
\sign([\R_{\hat{\mathcal{I}}}\A_{\hat{\mathcal{I}}}^H(\A_{\hat{\mathcal{I}}}\R_{\hat{\mathcal{I}}}^2\A_{\hat{\mathcal{I}}}^H + \sigma^2\A\G^{-1}\A^H)^{-1}\y]_n), & n\in\hat{\mathcal{I}};\\
0, & n\notin\hat{\mathcal{I}}.
\end{array}\right. \label{MMSE_RDMUD}
\end{equation}

\subsection{Maximum likelihood detector}

The optimal detector that minimizes the probability-of-error for the RD-MUD output is the nonlinear maximum likelihood detector. The maximum likelihood detector finds the active users and symbols by minimizing the likelihood function, or equivalently, minimizing the quadratic function $\|{(\A\G^{-1}\A^H)^{-1/2}}(\yb - \A\R\bb)\|^2$. This is also equivalent to solving the following integer optimization problem
\begin{equation}
\begin{split}
\max_{b_n\in\{-1, 0, 1\}
}
&2\yb^H (\A\G^{-1}\A^H)^{-1}\A\R\bb\\
&- \bb^H \R\A^H(\A\G^{-1}\A^H)^{-1}\A\R\bb, \label{ML_RDMUD}
\end{split}
\end{equation}
where $b_n = 0$ corresponds to the $n$th user being inactive.

\subsection{Choice of $\A$}
The coefficient matrix $\A$ is our design parameter. In Section \ref{sec:RDD} and Section \ref{sec:algorithm_II} we have shown that both the RDD and  RDDF detectors are based on the inner products between the projected received signal vector and the columns of $\A$. Hence, intuitively, for the RDD and RDDF detectors  to work well, the inner products between columns of $\A$, or its coherence defined in (\ref{def_coherence}) should be small. In the following we consider the random partial discrete Fourier transform (DFT) matrix, whose coherence is small and it is formed by randomly selecting rows of a DFT matrix $\textbf{F}$: $[\textbf{F}]_{nm} = e^{i\frac{2\pi}{N}nm}$ and normalizing the columns of the sub-matrix, where $i = \sqrt{-1}$.

\section{Performance of RD-MUD}\label{sec:RD-MUD-performance}

In the following, we present conditions under which the RDD and RDDF detectors can successfully recover active users and their symbols. The conditions depend on $\A$ through its coherence and are parameterized by the crosscorrelations of the signature waveform through the properties of the matrix $\G$. Our performance measure is the probability-of-error, which is defined as the chance of the event that the set of active users is detected incorrectly, \textit{or} any of their symbols are detected incorrectly:
\begin{equation}
P_{e} = 
P(\hat{\mathcal{I}}\neq \mathcal{I})+P(\{\hat{\mathcal{I}}= \mathcal{I}\}\cap \{\hat{\bb}\neq\bb\}).\label{Pe}
\end{equation}
We will show that the second term of (\ref{Pe}) is dominated by the first term. The noise plays two roles in the $P_e$ of (\ref{Pe}). First, the noise can be sufficiently large relative to the weakest signal such that a nonactive user is determined as active; second, the noise can be sufficiently large such that the transmitted symbol plus noise is detected in an incorrect decision region and hence decoded in error. 
The first error term in (\ref{Pe}) is related to the probability-of-error for support recovery (see, e.g. \cite{FletcherRanganGoyal2010}). There are two major differences in our results on this aspect of RD-MUD performance relative to those previous works. First, although noise in the analog signal model (\ref{sig_model}) is white, matched filtering at the RD-MUD front-end introduces colored noise in (\ref{RD_MUD_model}).
Second, we take into account the second term in (\ref{Pe}),  which has not been considered in previous work. 

Define the largest and smallest channel gains as
\begin{equation}
|r_{\max}|\triangleq \max_{n=1}^N  |r_n|,\quad |r_{\min}|\triangleq \min_{n=1}^N |r_n|. \label{gain_def}
\end{equation}
Our main result is the following theorem:
\begin{thm}\label{thm_noisy}
Let $\bb \in \mathbb{R}^{N\times 1}$ be an unknown deterministic symbol, $b_n \in \{-1, 1\}$, $n\in\mathcal{I}$, and $b_n = 0$, $n\in\mathcal{I}^c$, $n = 1, \cdots, N$. Assume that the number of active users $K$ is known. Given the RD-MUD front-end output $\y = \A \R \bb + \wb$, where $\A \in \mathbb{C}^{M\times N}$ and $\G \in \mathbb{R}^{N \times N}$ are known, and $\wb$ is a Gaussian random vector with zero mean and covariance $\sigma^2 \A\G^{-1}\A^H$, if the columns of $\A$ are linearly independent and the coherence of $\A$ (\ref{def_coherence}) satisfies the following condition: 
\begin{equation}
|r_{\min}| - (2K - 1)\mu |r_{\max}| \geq 2\tau,
\label{cond_thresholding}
\end{equation}
for some constant $\alpha >0$, and $N^{-(1+\alpha)} [\pi(1+\alpha)\log N]^{-1/2} \leq 1$, where 
\begin{equation}
\tau\triangleq \sigma \sqrt{2 (1+\alpha)\log N} \cdot \sqrt{\lambda_{\max} (\G^{-1})} \cdot\sqrt{\max_n \left(\ab_n^H\A\A^H\ab_n\right)}, \label{def_tau}
\end{equation}
then the probability-of-error (\ref{Pe}) for the 
RDD detector is upper bounded as:
\begin{equation}
P_e \leq N^{-\alpha} [\pi(1+\alpha)\log N]^{-1/2}.\label{high_prob_noisy}
\end{equation}
If the columns of $\A$ are linearly independent and the coherence of $\A$ (\ref{def_coherence}) satisfies a weaker condition:
\begin{equation}
|r_{\min}| - (2K - 1)\mu |r_{\min}| \geq 2\tau, \label{cond_OMP}
\end{equation}
for some constant $\alpha >0$, and $N^{-(1+\alpha)} [\pi(1+\alpha)\log N]^{-1/2} \leq 1$, then the probability-of-error (\ref{Pe}) for the RDDF detector is upper bounded by the right hand side of (\ref{high_prob_noisy}).
\end{thm}
The proof for Theorem \ref{thm_noisy} is given in \cite{XieEldarGoldsmithJournal2011}. The key idea of the proof is to find a uniform bound for the tail probability of the correlator output noise. Note in Theorem \ref{thm_noisy} that the condition of having small probability-of-error for the RDDF detector is weaker than for the RDD detector.
Based on the coherence of the random partial DFT matrix, we can prove the following corollary to Theorem \ref{thm_noisy} (see \cite{XieEldarGoldsmithJournal2011} for more details):
\begin{corollary}\label{DFT_logN}
Consider the setting of Theorem \ref{thm_noisy}, 
where $\A$ is a random partial DFT matrix $\A$. Suppose the number of correlators satisfies the following lower bound for the RDD detector
\begin{equation}
M\geq 4\left[\frac{(2K-1)|r_{\max}|}{|r_{\min}| -  2\tau} \right]^2 (2\log N + c),
\end{equation}
or satisfies the following smaller lower bound for the RDDF detector 
\begin{equation}
M\geq 4\left[\frac{(2K-1)|r_{\min}|}{|r_{\min}| -  2\tau} \right]^2 (2\log N + c),
\end{equation}
for some constants $c>0$ and $\alpha > 0$, and $|r_{\min}| >  2\tau$ for $\tau$ defined in (\ref{def_tau}), then the probability-of-error $P_e$ of the RDD detector or the RDDF detector is bounded by
\begin{equation}
1-(1-N^{-\alpha}[\pi(1+\alpha)\log N]^{-1/2})(1-2e^{-c}),
\end{equation}
for some constant $\alpha > 0$.
\end{corollary}
This corollary says that to attain a small probability-of-error, the number of correlators needed by the RDD and RDDF detectors is on the order of $\log N$, which is much smaller than that required by the conventional MUD using the MF-bank, which is on the order of $N$.

\section{Numerical Examples}\label{sec:numerical_eg}

\begin{figure}[h]
\centering{
\includegraphics[width = 2in]{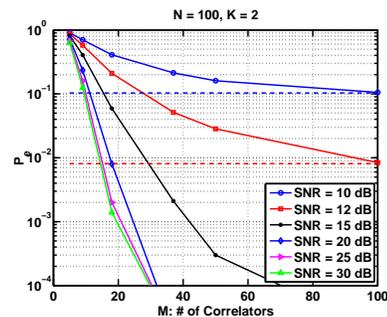}}
\vspace{0.1in}
\caption{Performance of the RDD detector, $P_e$ versus $M$ for different SNRs, when the signature waveforms are orthogonal, i.e., $\G=\I$.  The dashed lines show $P_e$ for the conventional decorrelating detectors at the corresponding SNR. When SNR is greater than 15 dB, (with $N = 100$ correlators) the probability-of-error of the decorrelating detector is less than $10^{-4}$. }\label{Fig:Pe_SNR}
\end{figure}

As an illustration of the performance of RD-MUD, we present an numerical example using the RDD detector. The results are obtained from $10^5$ Monte Carlo trials. For each trial, we generate a Gaussian random noise vector $\wb$ as well as a random partial DFT matrix for $\A$, and form the signal vector according to (\ref{RD_MUD_model}). To simplify, we assume that the gains for all the users are the same: $|r_{\min}| = |r_{\max}| =1$. Assume the signature waveforms are orthogonal ($\G = \I$). In this case, the noise in (\ref{RD_MUD_model}) is white. We define the signal-to-noise-ratio (SNR) as $|r_{\min}|^2/\sigma^2 = 1/\sigma^2$. We also assume $N = 100$ and $K = 2$. 
Fig. \ref{Fig:Pe_SNR} shows $P_e$ versus $M$ for the RDD detector as SNR increases. The counterpart of RD-MUD with the RDD detector in the conventional MUD setting is the decorrelating detector (when no subspace projection happens, i.e., if we let $\A = \I$ in (\ref{RD_MUD_model})). For each SNR, as $M$ increases,  the $P_e$ of the RDD detector approximates that of the conventional decorrelating detector. 
Also with higher SNR, the $P_e$ of the RDD detector decreases faster with increasing $M$. When SNR is sufficiently high, the number of correlators required by the RDD detector to achieve a small $P_e$ is much fewer than $N$.

\vspace{0.2in}

\vspace{-0.1in}

\section{Conclusions}\label{sec:conclusion}

We have developed families of digital detectors for the reduced-dimension multiuser detection (RD-MUD), and proved performance guarantees for two specific detectors: the reduced-dimension decorrelating (RDD) detector and the reduced-dimension decision feedback (RDDF) detector. This method reduces the number of correlators at the front-end of a MUD receiver by exploiting the fact that the number of active users is typically much smaller than the total number of users in the system. Motivated by the idea of analog compressed sensing, the RD-MUD front-end projects the received signal onto a lower dimensional detection subspace by correlating with a set of correlating signals. We proved that when the random partial DFT matrix is used to construct correlating signals for RD-MUD, the number of correlators is on the order of log of the number of users in the system, which is much smaller than that required by the conventional MUD. Numerical examples validated our theoretical results.

\bibliography{yao}

\end{document}